\documentstyle[11pt,newpasp_charb,twoside]{article}
\def\edcomment#1{\iffalse\marginpar{\raggedright\sl#1\/}\else\relax\fi}
\reversemarginpar

\begin{document}
\title{A Review of the Current Status of Follow-Up Techniques to Study Known Extrasolar Planets}
\author{David Charbonneau}
\affil{California Institute of Technology, MC 105-24 (Astronomy),\\1200 E. California Blvd., Pasadena, CA 91125 USA; dc@caltech.edu}

\begin{abstract}
I present a review of observational efforts to
study known extrasolar planets by methods that are
complementary to the radial velocity technique.  I describe
the current state of attempts to detect and
characterize such planets by astrometry, by reflected light, by thermal emission, 
by transit photometry, by atmospheric transmission spectroscopy,
by planet-induced chromospheric activity, and by
long-wavelength radio emission.  With a few exceptions,
these efforts have yielded only upper limits.
Nonetheless, the diversity and vivacity of these pursuits
has rapidly pushed many of these techniques into the realm
where realistic models of the planets and their atmospheres
can now be confronted.
\end{abstract}

\section{Introduction}
Doppler surveys of nearby Sun-like stars continue to supply
the bulk of information by which we are learning about
the planets that orbit our stellar neighbors.  The current
state-of-the-art in radial velocity precision did not
come easily:  It is the result 
of years of dedicated effort by numerous groups to develop 
the optimal instrumental methods, prior to the announcement 
of the first successful detections.  By analogy, dozens of 
research groups are
now exploring novel techniques to follow-up and characterize 
these newly-detected planets.  Many of these methods 
were considered completely unrealistic prior to
discovery of the great diversity of planetary systems,
when only our own Solar system served as a guide.
Notably, several techniques have been fueled by the 
detection of gas giants that are either significantly
more massive, or closer to the central star, than any object
in our Solar system.  I describe
here the current sensitivity of several of these techniques,
discussing wherever possible published data, and the
corresponding upper limits and/or detections.  I provide an 
extensive reference list so that this paper may also serve
as gateway for individuals new to the field:  Such studies
will likely require many teams and years of development to find 
the optimal observing strategy.  

\section{Astrometry}
Benedict et al. (2002) report the first astrometrically determined
mass of an extrasolar planet:  Capitalizing on the high-precision 
astrometry enabled by the HST Fine Guidance Sensors, they 
have detected the astrometric motion of the GJ~876 due to 
its outer orbiting planet (GJ~876~b, $P = 61.0$~d;
the inner planet, GJ~876~c, has a period of $P = 30.1$~d).  Combining
the astrometry with the radial velocity measurements, they solve for the
perturbation semi-major axis $\alpha = 0.25 \pm 0.06$~mas, inclination $i = 84\deg
\pm 6\deg$, as well as the GJ~876 absolute parallax, 
$\pi_{\rm abs} = 214.6 \pm 0.2$~mas.  Since the system is very close to
edge on, the planetary mass they determine, $M_{p} = 1.89 \pm 0.34\ M_{\rm Jup}$,
is indeed the minimum mass derived from radial velocities alone.
The high inclination indicates that the one or both
of the planets may indeed lie in a transiting configuration.  Furthermore,
the dynamical interaction of the two planets causes the orbital
elements to vary in time, indicating that one of the planets may
enter a transiting configuration, even if this is not currently
the case.

By combining Hipparcos astrometry with
radial velocity measurements, Han, Black, \& Gatewood (2001) argue
that at least 4 of the 30 companions with planetary
minimum masses that they examined possess 
stellar mass ($M > 80\ M_{\rm Jup}$), implying 
that these orbits are seen nearly face-on ($i < 1\deg$).
Pourbaix \& Arenou (2001) present a detailed discussion
demonstrating that these conclusions are likely an artifact of the
reduction procedure as applied to the Hipparcos data.
Astrometry of several additional extrasolar planet systems are presented 
in detail in the literature:  Gatewood, Han, \& Black (2001) claim
an inclination of $i = 0.5\deg$ and a mass of $M = 115\ M_{\rm Jup}$ 
for the companion to $\rho$~CrB (which has a minimum mass 
of $M \sin{i} = 1.04\ M_{\rm Jup}$), and Zucker \& Mazeh (2000) find
a true mass of $M = 38\ M_{\rm Jup}$ for the companion to
HD~10697 (which has a minimum mass of $M_{p} \sin{i} = 6.4\ M_{\rm Jup}$), 
indicating a brown dwarf companion rather than a planet.  McGrath et al. (2002, 2003)
have used HST/FGS to constrain the astrometric motion due to the
companion to $\rho^{1}$~Cnc to less than 0.3~mas (5~$\sigma$),
ruling out the preliminary Hipparcos value of 1.15~mas.  They derive
an upper limit on the companion mass of $30\ M_{\rm Jup}$, 
confirming that the object is substellar.

Currently, Benedict et al. (2003) are using HST/FGS to monitor 
the astrometric motion of both $\epsilon$~Eri and $\upsilon$~And.
They plan to monitor these systems for 3 years, but based on the
high precision (1 mas) of the data gathered so far (covering baselines
of 2 years and 1 year, respectively), they can already offer 
interesting upper limits on the companion masses.

\section{Reflected Light: Direct Spectroscopic Separation}
Planets shine in reflected light with a flux (relative to their stars) of
\begin{equation}
\left( \frac{f_{p}}{f_{*}} \right)_{\lambda} (\alpha) = \left( \frac{R_p}{a} \right )^2 \, p_{\lambda} \, \Phi_{\lambda}(\alpha),
\end{equation}
where $a$ is the semi-major axis, $p_{\lambda}$ is the
geometric albedo, and $\Phi_{\lambda}(\alpha)$ is the phase
function (the flux from the planet when viewed at a phase
angle $\alpha$ relative to the flux received when the
planet is at opposition).  The system can thus be pictured as an
extreme double-lined spectroscopic binary:  The secondary spectrum
is exceptionally faint ($f_{p}/f_{*} \sim 1 \times 10^{-4}$), but
very well separated spectroscopically, as the orbital velocities
for these hot Jupiters are typically of order $100\ \rm{km\, s^{-1}}$, 
much greater than the typical stellar line widths of $2 - 10\ \rm{km\, s^{-1}}$. 
The challenge is to detect the secondary spectrum, and thus derive
the planetary mass (from the projected velocity separation of the planet
and star), as well as estimates of the geometric albedo and planetary radius.

Charbonneau et al. (1999) placed the first significant upper limit on the reflected 
light from an extrasolar planet.  They observed $\tau$~Boo (orbital period
$P = 3.31$~d, $M_{p} \sin{i} = 4.38~M_{\rm Jup}$) with Keck/HIRES for 
3~nights under intermittent clouds.  
Assuming a planetary radius of $R_{p} = 1.2~R_{\rm Jup}$, 
they placed an upper limit on the average planetary geometric albedo 
of $p < 0.3$ over the wavelength range $465 - 499$~nm.  This limit is
valid for high orbital inclinations ($i > 70{\deg}$), and a less restrictive 
limit applies for lower inclinations.  Recently,
Leigh et al. (2003b) have presented a combined analysis of their WHT/UES data
of the $\tau$~Boo system from 1998, 1999, \& 2000.  Assuming a planetary radius
of $R_{p} = 1.2~R_{\rm Jup}$, an orbital inclination of $i = 36{\deg}$, 
and a Venus-like phase function, they present an 
upper limit on the average geometric albedo 
(taken over their wavelength range of $385 - 611$~nm) of 
$p < 0.39$ (for a conservative false alarm 
probability of 0.1\%).  This system is distinctive in that there
is independent evidence that the star is tidally locked at the 3.31~d orbital period 
(Baliunas et al. 1997).  As a result, the reflected light spectrum is likely
non-rotationally broadened (with a line width determined by the macroturbulent 
velocity, $v_{turb} \simeq 4~{\rm km\, s^{-1}}$), which should facilitate its
separation from the rotationally-broadened 
stellar features ($v_{*} \sin{i} \simeq 15~{\rm km\, s^{-1}}$).

Similar studies have been carried out for two other hot Jupiter systems:
Based on 3 nights of WHT/UES spectra, Collier Cameron et al. (2002)
derived an upper limit for the geometric albedo of $\upsilon$~And~b
($P = 4.62$~d, $M_{p} \sin{i} = 0.73~M_{\rm Jup}$) of 
$p < 0.98~(R_{p}/R_{\rm Jup})^{-2}$ (0.1\% false alarm probability).
They consider two model spectra from Sudarsky et al. (2000), and
place limits on the planetary radius based on these models.  The
first is a Class V ``Roaster'' model atmosphere that includes
the effects of stellar insolation.  This model yields a predicted
geometric albedo of $p = 0.42$ when averaged over their band pass,
and thus requires that $R_{p} < 1.51~R_{\rm Jup}$.  The second model
they consider is an isolated Class IV model atmosphere, which pedicts 
a lower geometric albedo of $p = 0.19$, and yields a correspondingly 
weaker limit on the planetary radius of  $R_{p} < 2.23~R_{\rm Jup}$.  
Using 4 nights of 
VLT/UVES data, Leigh et al. (2003c) have presented
a similar study of the hot Jupiter system HD~75289 
($P = 3.51$~d, $M_{p} \sin{i} = 0.46~M_{\rm Jup}$).  Assuming an
orbital inclination of $i = 60{\deg}$ and a planetary radius of
$R_{p} = 1.6~R_{\rm Jup}$, they placed an upper limit on the average
geometric albedo of $p < 0.12$, with a false alarm probability of 0.1\%.

Charbonneau \& Noyes (2000) and, more recently, Leigh et al. (2003a) have
discussed future prospects for reflected-light studies.  Two attractive
targets for which no such studies have been published to date are
51~Peg ($P = 4.23$~d, $M_{p} \sin{i} = 0.48~M_{\rm Jup}$) and
HD~179949 ($P = 3.09$~d, $M_{p} \sin{i} = 0.84~M_{\rm Jup}$).  The HD~209458
system ($P = 3.52$~d, $M_{p} \sin{i} = 0.69~M_{\rm Jup}$) would
require longer integration times due to its faintness ($V = 7.6$), yet the
interpretation of such data would be greatly simplified:  Since the
orbital inclination ($i = 86.7{\deg}$) and planetary radius 
($R_{p} = 1.35~R_{\rm Jup}$) are known, the geometric albedo can be 
estimated unambiguously from the flux ratio.

\section{Reflected Light: Photometric Monitoring}
A more direct method to study the albedo and scattered light curves
of hot Jupiters is ultra-high precision photometry capable of
measuring the photometric modulation that results from the
varying illumination of the planet as it orbits the central star.
Ground-based efforts to detect the reflected-light
modulation from extrasolar planets are typically limited by
atmospheric extinction effects to a precision of 0.1\%
(see Kenworthy \& Hinz 2003 for a description of such
an effort).  Thus it appears that the required precision
will be available only from space for the time being.
The MOST satellite (Matthews et al. 2000; Walker et al. 2003) was 
launched 30 June 2003, and possesses just such a capability:  
Although its primary science goal is the detection and characterization of 
rapid acoustic oscillations in Sun-like stars ($\nu \simeq 0.5 - 6$~mHz), 
it should have the precision to detect the variation caused by several known 
hot Jupiters ($\nu \simeq 0.003$~mHz).  MOST conducts photometry
in a single bandpass ($350 - 700$~nm) and thus will provide no spectral
information.  Nonetheless, the derived phase variation should be diagnostic of
the principal sources of scattering in the planetary atmosphere.
Three known hot-Jupiter systems, 51~Peg, $\tau$~Boo, and HD~209458, are
on the MOST primary target list.  Recently, Green et al. (2003) have presented 
a detailed analysis of the ability of MOST to measure such variations.
Assuming a conservative detection threshold of 4.2~$\mu$mag (3~$\sigma$),
they find that MOST should indeed be able to detect $\tau$~Boo~b in reflected
light, or place severe constraints on the models of the
planet atmosphere.

Jenkins \& Doyle (2003) present a detailed analysis of the ability 
of the upcoming Kepler Mission to detect previously unknown hot Jupiters
by their photometric variation.  They find that, depending on the
albedo and scattering function, Kepler should reveal $100 - 760$
new hot Jupiters (with orbital periods less than 7~days) by their reflected light,
assuming a conservative limit of not more than one false positive rate
over the Kepler campaign.  The primary detectability determinant
(other than the model of the planetary atmosphere) is the stellar
rotation period, although stellar brightness is significant for
$R > 12.5$.  The upcoming CNES COROT Mission (Schneider et al. 1998) may
also detect hot Jupiters by this effect.

\section{Transit Photometry}
I do not discuss here the numerous efforts to detect unknown
extrasolar planets by the transit method (for a review,
see Horne 2003 and Charbonneau 2003a).  Rather, I restrict my 
attention to the two systems that are known to transit.

Numerous groups (Charbonneau et al. 2000; Henry et al. 2000;
Jha et al. 2000; Deeg, Garrido, \& Claret 2001) have presented 
ground-based photometry of the planetary transit of HD~209458 with a typical
precision of 2~mmag and a cadence of roughly 10~minutes.  At this
level of measurement, estimation of the planet radius is frustrated
by a significant degeneracy, as it is possible to fit the data
by a family of models wherein the ratio $R_{p}/R_{*}$ is preserved.
The typical resulting uncertainty in the planetary radius
as derived from these ground-based data sets is 10\%.  In contrast
to this situation, Brown et al. (2001) observed four transits 
with HST/STIS, and achieved a precision of 0.1~mmag and a rapid 
cadence of 80~s.  These data constrained the slope and duration of 
ingress and egress with sufficient quality to break the above degeneracy.
The resulting estimates of the planetary
and stellar radii were $R_{p} = 1.35 \pm 0.06\ R_{\rm Jup}$ and $R_{*} = 
1.15 \pm 0.05\ R_{\sun}$.  Combining the planetary radius with
the planetary mass $M_{p} = 0.69 \pm 0.05\ M_{\rm Jup}$, 
Brown et al. (2001) deduce an average density of $\rho = 0.35\ {\rm g\, cm^{-3}}$, 
a surface gravity of $g = 943\ {\rm cm\, s^{-2}}$, 
and an escape velocity of $v_{e} = 43\ {\rm km\, s^{-1}}$.  
They also ruled out the presence
of planetary satellites larger than 1.2~$R_{\earth}$ (based
on photometric residuals) or more massive
than 3~$R_{\earth}$ (based on timing residuals), and
placed constraints on the presence of opaque circumplanetary rings.
Schultz et al. (2003) observed several transits with
the HST Fine Guidance Sensors, which yielded an ultra-rapid cadence
of 0.025~s, and a typical signal-to-noise ratio of 80.  These
observations target times of ingress and egress, and can be
used to search for timing offsets indicative of planetary satellites.
Furthermore, the combination of the Brown et al. (2001) and
Schultz et al. (2003) data sets, each of which permit an 
accurate determination of the time of center of transit, yields
an extremely precise value of the orbital
period.  This should effectively remove any timing uncertainties
in the planning and interpretation of future data sets.

OGLE-TR-56~b is the first extrasolar planet detected by the transit
method (Udalski et al. 2002a,b; Konacki et al. 2003a).  The period
deduced from recent Doppler observations of the system (Torres et al. 2003) 
confirms that found from the photometric data, and the amplitude of
the Doppler shift demonstrates that the companion is indeed of planetary mass.  
The precision of the radial velocity 
observations is relatively coarse (typically 100~m$\,{\rm s}^{-1}$, 
due to the much fainter magnitude of the star than those typically
surveyed).  The best estimates of the planetary radius, 
$R_{p} = 1.23 \pm 0.16\ R_{\rm Jup}$, and mass,
$M_{p} = 1.45 \pm 0.23\ M_{\rm Jup}$, imply a much greater
density  $\rho = 1.0\ {\rm g\, cm^{-3}}$ than that of
HD~209458~b.  The semi-major axis
of OGLE-TR-56~b (0.0225~AU) is less than half that of HD~209458~b (0.0468~AU),
and thus comparison of these objects may offer us insight
into the importance of insolation upon the planetary radius.
Several other targets from the OGLE campaign remain
under scrutiny (Konacki et al. 2003b).

\section{Transmission Spectroscopy of the Atmosphere and Exosphere}
Charbonneau et al. (2002) observed four transits of HD~209458 with 
HST/STIS, and detected an increase in
transit depth of $(2.32 \pm 0.57) \times 10^{-4}$ in a narrow
bandpass centered on the sodium resonance lines near 589.3~nm.
They rule out alternate explanations of this dimming (such
as stellar limb darkening), and
conclude that the effect is due to absorption in the planetary atmosphere.
The signal amplitude is roughly 1/3 that predicted from model
atmospheres that are cloudless and contain a solar abundance of
sodium in atomic form.  A possible explanation for this
decrement is the presence of a high-altitude cloud deck (with cloud
tops above 0.4~mbar).  An alternative reason is the depletion
of atomic sodium, either as a result of chemical reaction into
molecules such as NaCl and Na$_2$S, or indicative of a 
global metal abundance that is significantly reduced relative to solar.

Similarly, Vidal-Madjar et al. (2003) observed three transits of HD~209458
with HST/STIS, but in a bandpass centered on the Ly$\alpha$ feature
(notably, these UV wavelengths use the STIS MAMA detectors,
as opposed to the STIS CCD device).  They found a transit depth of
$15 \pm 4$~\%, corresponding to an equivalent size of 4.3~$R_{\rm Jup}$.
As this is in excess of the Roche limit of 3.6~$R_{\rm Jup}$, they
conclude that some hydrogen atoms must be escaping from the
planet (there is also some independent spectral evidence that 
the atoms have a large velocity relative to the planet).  Although
the minimum escape rate required by the data would
reduce the planetary mass by only a negligible amount (0.1\%)
when integrated over the age of the system,
significantly higher mass-loss rates are also permitted.

There have been several ground-based efforts to detect additional
atmospheric features in absorption, all of which have been
frustrated by variability in the telluric spectrum and the spectrograph.
Moutou et al. (2001) observed a portion of a planetary transit with VLT/UVES,
spanning the wavelength range $328 - 669$~nm, and performed a general
search for prominent absorption features.  Their precision was typically
1\% (worse at shorter wavelengths), and thus they were restricted to
looking for exospheric (as opposed to atmospheric) features.  Such
an exosphere is undoubtedly the site of complex photochemistry, which
results in a long list of potential atomic and molecular species (and their 
associated ions).  Moutou et al. (2003a) observed a planet transit with
the VLT/ISAAC instrument to search for the He~{\scriptsize I} line at
1083~nm.  They present an upper limit of 0.5\% (3~$\sigma$) for
a 0.3~nm band centered on the feature, slightly above the predicted
signal strength of 0.25\%.  They outline improvements that would
likely allow similar observations to achieve a detection limit of 0.1\%.
Iro et al. (2003) \& Moutou et al. (2003b) have also presented 
preliminary results from associated efforts.  

Brown, Libbrecht \& Charbonneau (2002) present Keck/NIRSPEC
data from a preliminary attempt to detect features from the 
CO molecule at 2.3~$\mu$m.  The CO molecule is diagnostic
of the temperature in the planetary atmosphere, as the equilibrium
temperature of HD~209458b ($\sim$1400~K) falls in the regime where
carbon migrates from CO to CH$_4$ (with the latter preferred
at lower temperatures).  The observing conditions, however, were not ideal:
The transit was not visible in its entirety from their longitude,
and the weather was poor.  They present an upper limit that is
a factor of 3 above reasonable models of the planetary atmosphere,
and argue that future attempts should be able to achieve
the precision required to test directly models of the planetary atmosphere.
Harrington et al. (2003) discuss a large effort to detect IR
transmission features with data gathered with Palomar, Keck, VLT, and IRTF.

Rauer et al. (2000) present spectra of 51~Peg gathered with the SWS 
instrument on ISO.  Their sensitivity is sufficient to
place limits on the presence of features from an extended exosphere.
However, the lack of constraints on the orbital inclination prevent
them from directly constraining models of the planetary exosphere.

\section{Infrared Emission}
Infrared wavelengths offer a far more favorable contrast
ratio between the planet and star than visible light.  As
a result, there
have been numerous efforts to detect directly IR spectroscopic
features from the planetary atmosphere.  Lucas \& Roche (2002) 
examined five systems (HD~187123, 51~Peg, $\tau$~Boo, $\upsilon$~And,
and HD~209458) with the UKIRT Cooled Grating Spectrograph 
CGS4 at $K$ and $L'$.  They searched for the spectroscopic edges 
arising from H$_2$O and CH$_4$, neither of which are
expected to be present in the stellar spectrum.  Their typical
3$\sigma$ upper limits for water features were at the level of one part in several
hundred.  The corresponding limits on the planetary radius ranged
from $2.6 - 5~R_{\rm Jup}$ (depending upon the individual object
and atmospheric model) and hence were insufficient to
test directly realistic models of the planetary radii and atmospheres.
Nonetheless, they note that similar studies with SIRTF would likely
attain the required precision.  Wiedemann, Deming, \& Bjoraker (2001)
present the results of a sensitive search for methane
in the infrared spectrum of $\tau$~Boo, using the IRTF CHSELL spectrometer.
In contrast to Lucas \& Roche (2002), they explicitly search
for a signal that varies according to the known planetary orbit.
They tentatively claim a detection of methane features in
the planetary atmosphere (the signal strength is $2 \times 10^{-4}$
of the stellar continuum), but they caution that the significance
of the detection is marginal (2.4~$\sigma$).

In a couple of recent papers, Richardson et al. (2003a,b) have 
presented strong limits on the infrared spectrum of HD~209458~b.
Their method of ``occultation spectroscopy'' capitalizes 
on the transiting geometry to search for the disappearance and 
reappearance of spectral features at the times of ingress and egress.
They argue that the signal amplitude available to this method exceeds
the analogous signal from transmission spectroscopy for wavelengths
longward of approximately 2.6~$\mu$m.  In Richardson et al. (2003a),
they present the analysis of VLT/ISAAC spectra with a resolution
of $R = 3300$ gathered during two times of secondary eclipse.
They search for planetary methane features near 3.6~$\mu$m,
and are able to rule out irradiated, low-opacity (cloudless),
low-albedo, thermochemical-equilibrium models of the planetary
atmosphere.  They are not, however, sensitive to all reasonable
models of the planetary atmosphere:  In particular, cloudy models
(as favored by the small amplitude of the sodium feature detected
by transmission spectroscopy; Charbonneau et al. 2002) would
yield substantially weaker features which would be below their
detection threshold.  Richardson et al. (2003b) present data
spanning the $1.9 - 4.2\ \mu$m region gathered with the IRTF/SpeX
instrument with a resolution of $R = 1500$.  They give upper
limits on a continuum peak near 2.2$\mu$m that would result from
CO and H$_2$O features in the planetary atmosphere.  These upper
limits are at the level of roughly $3 \times 10^{-4}$
of the stellar flux, a factor of two below some models of
the planetary atmosphere.  They are able to rule out cloudless
models in which there is a strong day-night temperature asymmetry,
as such models yield a steep temperature-pressure profile to which
their experiment is most sensitive.

The secondary eclipse need not occur exactly half a period after
the primary eclipse.  The current upper limit on the orbital
eccentricity, $e = 0.00967 \pm 0.014$ (G. Marcy, personal 
communication) is indeed consistent with zero (as expected from
tidal circularization), but could conceivably be as high as $\sim 0.04$.
An eccentricity of $e \simeq 0.04$ could shift the time of secondary
eclipse by as much as 2~hours (depending on the longitude of
periastron $\omega$; see Charbonneau 2003b for details).  The detection 
of a significant offset would place a lower limit on the eccentricity (and the radial 
velocity measurements provide an upper bound).  Conversely, 
upper limits on the offset would provide even more stringent constraints
upon the eccentricity, and thus bear upon the mechanisms that 
compete to pump and damp the orbital eccentricity.
Thus, experiments that attempt to detect the secondary eclipse
must account for the additional complication that the event may
shifted by as much as 2~hours, but this possibility also provides
us with a additional tool with which to study the planet and
its orbit.

\section{Planet-Induced Chromospheric Activity}
Cuntz, Saar \& Musielak (2000) investigate several possible interactions
between extrasolar planets and their parent stars that may result in 
stellar chromospheric and coronal activity enhancement.  These effects
will be due predominantly to a combination of tidal and magnetic effects.
The origin of any observed activity could be distinguished observationally,
since tidal effects should reveal a periodicity of $P/2$ (where $P$ is
the planetary orbital period), whereas magnetic effects should follow
a period of $P$.  Both tidal and magnetic effects would decrease rapidly
in intensity with semi-major axis $a$ (with dependencies of $\propto a^{-3}$ 
and $\propto a^{-2}$, respectively).  Thus the best targets are the hot Jupiter systems.
These planets may, however, have magnetic fields that are significantly
weaker than that of Jupiter, since they are likely tidally locked,
and thus their rotation periods are typically 10 times greater than the
Jovian value.

Saar \& Cuntz (2001) searched for periodicities in the Ca {\scriptsize II} infrared
triplet, indicative of chromospheric activity, in 7 extrasolar planet
systems ($\tau$~Boo, 51~Peg, $\upsilon$~And, $\rho^{1}$~Cnc, $\rho$~CrB,
70~Vir, and GJ~876).  They examined extant Lick spectra, with a typical
resolution $R = 50,000$ and a signal-to-noise ratio of 200, which were
gathered as part of the extrasolar planet search program.  
For each spectrum, they evaluate the index 
$S_{\rm IR} = \langle F_{\rm em} \rangle /F^{*}_{\rm con}$, where
$\langle F_{\rm em} \rangle$ is the emission flux centered on the core 
of the 866.2~nm line, and $F^{*}_{\rm con}$ is the median continuum flux 
in a line-free region on either side of the line.  They found no significant
signal at either period ($P/2$ or $P$), with upper limits of $3 - 10$~\%,
although for several stars they saw low-statistical-significance indications 
of the stellar rotational signature.

Shkolnik, Walker, \& Bohlender (2003a) have recently detected the enhancement
of Ca {\scriptsize II} H and K emission in phase with the orbit ($P = 3.09$~d)
of the hot Jupiter companion to HD~179949, indicating a magnetic (rather
than tidal) origin of the signal.  This detection may offer the 
community's first access to the magnetosphere of an extrasolar planet.  
For this effort, 
they gathered multi-epoch high-resolution spectra ($R = 110,000$) with a 
typical signal-to-noise of 500 (in the 
continuum adjacent to the Ca {\scriptsize II} H and K lines).  They
also monitored a strong photospheric Al {\scriptsize I} line, which
revealed no changes, indicating that the periodic variation did indeed
originate in the chromosphere.  As of yet, there is no independent
evidence that the stellar rotation period is the same as the orbital
period.  Shkolnik et al. (2003b) are monitoring 4 other hot Jupiter systems.
Three of these ($\tau$~Boo, $\upsilon$~And, and HD~209458) have revealed
significant nightly variations, but, as of yet, the variations do not conclusively 
show phase coherence with the planetary orbits.  The group is
monitoring 51~Peg, but has seen no clear evidence for variability.

\section{Radio Emission}
Even before the Doppler detections of numerous 
extrasolar planets, searches for long-wavelength radio emission 
from planets orbiting neighboring stars had been carried out 
(Wingleee, Dulk \& Bastian 1986).  Such searches are motivated
by the observation that the polar regions of all of the Solar
system gas giants are the sites of intense auroral-related radio
emission.  In particular, the coherent cyclotron radiation from
Jupiter often exceeds $10^{10}$~W.  It is variable in time by a factor 
of 1000, and the emission is correlated exponentially
with the input velocity and power of the solar wind.  Several
scaling laws have been proposed to relate the approximate radio power
for the planets of the solar system.
These ``Radiometric Bode's Laws'' suggest that the radiated
power scales directly with the planetary mass, and inversely
with the three-halves power of the semi-major axis, $\propto M_{p}\ a^{-3/2}$.
Since we now know of several nearby Sun-like stars that are
host to Jupiter-mass planets at a distance of 0.05~AU, 
the scaling laws from the Solar system suggest that the radio
emission from these objects could exceed Jupiter by
a factor of 1000 (see discussion in
Farrell et al. 2003 for more details).  Farrell et al. (1999), 
Bastian et al. (2000), and Zarka et al. (2001) discuss
the emission mechanisms in detail, and make predictions of
emitted radio power for several of the known extrasolar planets.

The detection of cyclotron radio emission from extrasolar
planets would be a valuable tool (e.g. Bastian et al. 2000):  
It would permit the characterization of the planetary magnetic field,
and the upper limit on the emission frequency 
would yield the maximum strength of the magnetic field.
The sense of circular polarization might indicate from
which pole the emission originated, and the elliptical
polarization might yield a limit on the plasma density
in the planetary magnetosphere.  Furthermore, the periodicity
of the radio emission would give a direct determination
of the rotational period (indeed, the rotation periods
for the solar system gas giants are determined by this
method, not by the observation of atmospheric features,
which rotate differentially).  Additional periodicities
in the emission might indicate the presence of
planetary satellites, again by analogy with the planets
of the Solar system. 

Farrell et al. (2003) identified the hot Jupiter system
$\tau$~Boo ($M_{p} = 4.38~M_{\rm Jup}$) as the optimal target, 
as the scaling favors massive planets at small orbital
separations.  At a distance
of 10~pc, the radio power from Jupiter would be approximately 0.023~mJy. 
Based on the scaling law, $\tau$~Boo~b could yield
\begin{eqnarray}
S_{p} & \simeq & \left( \frac{M_{p}}{M_{\rm Jup}} \right) 
\left( \frac{a_{p}}{a_{\rm Jup}} \right)^{-3/2} \left( \frac{d}{10{\rm pc}} 
\right)^{-2} S_{\rm Jup,10pc} \\
 & \simeq & 4.38\ (0.046/5.2)^{-3/2}\ (15.6/10)^{-2}\ 0.023~{\rm mJy}\\ 
 & \simeq & 50~{\rm mJy}.  
\end{eqnarray}
Observing at the Very Large Array (VLA) in May 2002, they placed
an upper limit of 120~mJy for such emission at 74~MHz.
Although this limit does not yet reach the predicted signal
level, it demonstrates that such searches are nearly within
reach of current instrumentation, and should be accessible
to the next-generation arrays.  Furthermore, such
emission is likely quite variable in time.  Thus 
it could be that the emission from $\tau$~Boo~b does occasionally
exceed the current capabilities of the VLA.

Bastian et al. (2000) have conducted VLA observations of seven
extrasolar planets (51~Peg, $\upsilon$~And, ${\rho}^1$~Cnc,
47~UMa, $\tau$~Boo, 70~Vir, and HD~114762) at 333~MHz and 1465~MHz, 
where the sensitivities are greater, but the predicted signals
are much weaker.  Their typical 1-$\sigma$ sensitivities were $1 - 10$~mJy
at 333~MHz, and $0.02 - 0.07$~mJy at 1465~MHz.  They state
upper limits for each system.  They also observed the 47~UMa system
($M_{p}\ \sin{i} = 2.54\ M_{\rm Jup}$, $P =1089$~d), at 
74~MHz with a sensitivity of 50~mJy, but did not detect a signal.
A weaker signal is expected for 47~UMa than for the hot Jupiter
system $\tau$~Boo, due to the much larger semi-major axis.

\end{document}